\documentclass[showpacs,nofootinbib,reprint,prd,aps]{revtex4-1}

\usepackage{amsmath,amssymb}
\usepackage{epsfig}  
\usepackage{graphicx}               
\usepackage{url}
\usepackage{hyperref}
\usepackage{bm}
\usepackage{tikz}
\usepackage{verbatim}
\usepackage[utf8]{inputenc}

\usepackage{pstricks}
\usepackage{color}
\usepackage{slashed}

\newcommand{\nc}{\newcommand}  

\nc{\beq}{\begin{equation}}  
\nc{\eeq}{\end{equation}}  
\nc{\beqa}{\begin{eqnarray}}  
\nc{\eeqa}{\end{eqnarray}}  
\nc{\bit}{\begin{itemize}}  
\nc{\eit}{\end{itemize}}

\begin{document}
\title{Beauty-full Tetraquarks}
\author{Yang Bai, Sida Lu, and James Osborne} 
\affiliation{Department of Physics, University of Wisconsin-Madison,
  Madison, Wisconsin 53706, USA}
\begin{abstract}
In this article we present a calculation of the $bb\bar{b}\bar{b}$ tetraquark ground-state energy using a diffusion Monte Carlo method to solve the non-relativistic many-body system. The static potential for the four quark system is chosen to align with the flux-tube picture of QCD. Using this approach, we find that the $0^{++}$ state has a mass of $18.69\pm 0.03~\mbox{GeV}$, which is around 100 MeV below twice the $\eta_b$ mass. This bound state can behave as a four-lepton resonance via its decay to $\Upsilon(1S) \Upsilon(1S)^* \rightarrow \ell^+ \ell^- \ell^+ \ell^-$.
\end{abstract}
\maketitle

\section{Introduction}
\label{sec:intro}
In the seminal work of Ref.~\cite{Appelquist:1974zd}, Applequist and Politzer interpreted heavy quark bound states as positronium-like atoms subject to non-relativistic quantum mechanics calculations. The spectroscopy of quarkonia can then be well understood by solving Schr\"odinger's equation based on the static potential between two heavy quarks~\cite{Eichten:1974af,Quigg:1979vr,Kwong:1987mj} mediated by the asymptotically-free quantum chromodynamic (QCD) interactions. Ever since, two-body heavy quark systems have been used to understand the long distance behavior of QCD.

Multi-quark states were first proposed in 1964 by Gell-Mann as an explanation for the observed spectrum of mesons and baryons~\cite{GellMann:1964nj}. This now universally-accepted picture of mesons (baryons) as two- (three-)quark states has been hugely successful, and suggests that the mass of ordinary matter can be explained by the binding energy between quarks. More recently, enormous progress has been made both theoretically and experimentally in our understanding of four-quark states containing at least one light quark~\cite{Brambilla:2004jw,Chen:2016qju}. On the other hand, four-quark states containing only heavy quarks have not been directly confirmed by experimental searches. With new data from CMS, ATLAS, and LHCb at the LHC, a multi-quark state containing only bottom and/or charm quarks is very likely to be tested. From the theoretical side, such heavy quark states provide a unique environment to examine the non-relativistic QCD effective potential of many-body systems. In this paper, we concentrate on a potential tetraquark state comprised of two $b$ and two $\bar{b}$ quarks.

In QED, the equivalent system is the di-positronium molecule (Ps$_2$), first postulated by Wheeler in 1946~\cite{Wheeler}. Using the variational method, the binding energy against dissociation into two positronium atoms is calculated to be 0.435 eV~\cite{Hylleraas:1947zza,Ho:1986zz,Frolov:1996kq}, which is around 3.2\% of the binding energy of two positronium atoms. However, it wasn't until 2007 that the use of positron traps and accumulators led to the experimental confirmation of the Ps$_2$ molecular state~\cite{Mills:exp}. For the neutral positronium atoms, the electric dipole-dipole interaction can be used to generate the splitting between the ground and excited states. After the standard quantum mechanical perturbative calculation, the additional binding energy can be interpreted in terms of the $R^{-6}$ London-Van der Waals force~\cite{London}.

For QCD, there is no equivalent chromo-electric dipole interaction for a color-neutral meson, but one still has the transition-dipole interaction from color-neutral state to color-octet state. Depending on the relation between the inter-meson distance and intra-meson binding energy, second-order perturbative calculations reveal a similar $R^{-6}$ London-Van der Waals force~\cite{Appelquist:1978rt,Brambilla:2015rqa} or an $R^{-7}$ Casimir-Polder force~\cite{Appelquist:1977es,Peskin:1979va,Bhanot:1979vb}. As emphasized in Ref.~\cite{Appelquist:1978rt}, the Van der Waals force arises at leading order in $\alpha_s$ and depends only on the geometric ratio of inter-meson and intra-meson binding energies. In the heavy quark limit where the QCD effective potential is Coulomb-like, one can use the Van der Waals force to calculate the additional inter-meson binding energy and obtain a ratio of ${\cal O}(1\%)$ of the total intra-meson binding energy, similar to the Ps$_2$ case. For bottom quarks with finite mass, one should also include a long-range linear contribution to the potential~\cite{Eichten:1978tg}.

Several methods have been proposed in the literature to calculate the energies of tetraquark systems comprised of purely heavy quarks. One could, for example, rely on the QCD sum rule method as in Ref.~\cite{Chen:2016jxd}. However, the small separation between quarks in the four-$b$ system requires the use of higher-derivative moments to reliably estimate the non-perturbative bound state physics, which in turn calls into question the convergence of the perturbative expansion for such moments. Even for the two-$b$ system, this approach can become less trustable than the similar charmonium calculations~\cite{Reinders:1984sr}. A separate approach is to treat the $bb(\bar{b}\bar{b})$ system as a composite diquark bound state and then calculate the inter-diquark binding energy~\cite{Berezhnoy:2011xn,Karliner:2016zzc}. However, this simplified approach turns out to be inadequate as the average distance between the two diquarks is comparable to the separation between their constituent quarks, necessitating a complete four-body calculation.

In this paper, we adopt a phenomenological potential with its parameters determined by fitting to the two-body $b \bar{b}$ spectrum and verified by lattice simulation. We then numerically solve the Schr\"odinger equation to obtain the ground state energy and approximate wave function for the four-$b$ tetraquark. After that, we discuss spin-dependent (SD) corrections and obtain a final estimate for the ground state mass.

\section{Cornell Potential, Flip-flop, and Butterfly}
\label{sec:SI}

For a two-body $q\bar{q}$ system, the Cornell potential $V(r) = -4 \alpha_s/(3\,r) + r/a^2$ has been widely used to understand bottomonia and charmonia spectroscopy~\cite{Quigg:1979vr,Kwong:1987mj}. The $-1/r$ Coulomb term is understood simply as the spin-independent contribution from one-gluon exchange. The $r/a^2$ linear term is the long-range contribution due to QCD confinement. When one extends this phenomenological potential to many-body system, all flux-tube configurations must be checked to minimize the static potential. For three-body systems like baryons, lattice QCD calculations predict that only two flux-tube configurations matter; which one to choose depends on whether an interior angle of the triangle formed by the three quarks is greater than 120$^\circ$~\cite{Carlson:1982xi}.

\begin{figure}[thb!]
\begin{center}
\includegraphics[scale=0.55]{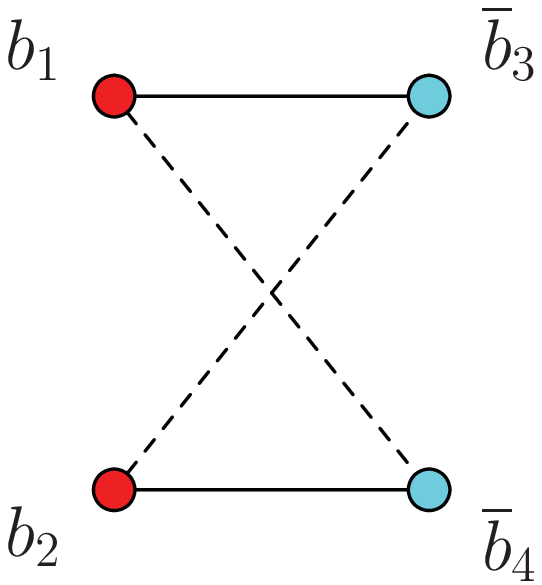} \hspace{8mm}
\includegraphics[scale=0.55]{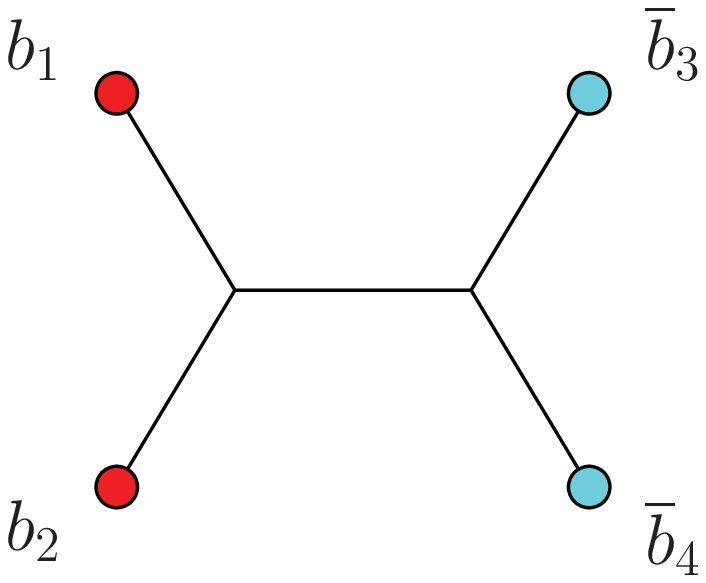}
\centering
\caption{Left panel: the flip-flop configuration of disconnected di-mesons. Right panel: the butterfly configuration with two connected diquarks. The two middle connecting points are chosen to minimize the total path. 
}
\label{fig:flip-flop-butterfly}
\end{center}
\end{figure}

For the four-body system there are in general three relevant configurations, with the flux-tubes reconfiguring in such a way as to minimize the total potential. The first two configurations, which in combination is sometimes referred to as the ``flip-flop'' configuration, is shown in the left panel of Fig.~\ref{fig:flip-flop-butterfly}. Depending on the relative distance between $(r_{13}, r_{24})$ and $(r_{14}, r_{23})$, one has the di-meson configuration of either $(b_1\overline{b}_3, b_2\overline{b}_4)$ or $(b_1\overline{b}_4, b_2\overline{b}_3)$. The third, so-called ``butterfly'' configuration, is shown in the right panel of Fig.~\ref{fig:flip-flop-butterfly}. This is the diquark-diquark configuration with two bottom quarks, $b_1$ and $b_2$, forming a color-anti-triplet (or less-favored sextet), and the two anti-bottom quarks, $\bar{b}_3$ and $\bar{b}_4$, forming a color-triplet. For the butterfly configuration, the middle two connecting points should be chosen to minimize the total flux-tube length.

The effective potential for the di-meson system is either
\beqa
V^{\rm di-meson}_{(13, 24)} = - \frac{4\,\alpha_s}{3} \left( \frac{1}{r_{13}} + \frac{1}{r_{24}} \right)  + \frac{1}{a^2}\left( r_{13} + r_{24} \right)
\eeqa
or $V^{\rm di-meson}_{(14,23)}$, which differs only by exchanging $\bar{b}_3 \leftrightarrow \bar{b}_4$. The flip-flop potential is defined as the minimum of the two di-meson configurations for any four-particle phase space and is given by
\beqa
V^{\rm flip-flop} \equiv \mbox{min}\left[ V^{\rm di-meson}_{(13, 24)}, V^{\rm di-meson}_{(14, 23)}\right] \,.
\eeqa
The effective potential for the butterfly configuration is
 \beqa
 V^{\rm butterfly} &=& - \frac{\alpha_s}{3} \left( \frac{1}{r_{13}} + \frac{1}{r_{14}} + \frac{1}{r_{23}} + \frac{1}{r_{24}}  \right)  \nonumber \\
 && -  \frac{2\,\alpha_s}{3}   \left( \frac{1}{r_{12}} + \frac{1}{r_{34}} \right) \,+\, \frac{1}{a^2}\,L_{\rm min}
 \,.
 \eeqa
 Here, $L_{\rm min}$ is the minimum value of the total flux-tube length for all possible connecting points in the right panel of Fig.~\ref{fig:flip-flop-butterfly}. The formula for the sextet diquark configuration is obtained by replacing the coefficients $(-1/3, -2/3)$ by $(-5/6, 1/3)$. As the sextet configuration has both attractive and repulsive Coulomb forces, it generically leads to a larger potential energy and thus will not contribute to the ground state energy calculation. The total four-quark potential is defined to be the minimum of the three possible configurations and is given by
\beqa
V^{4Q}  \equiv \mbox{min} \left( V^{\rm flip-flop},  V^{\rm butterfly} \right) \,. 
\eeqa

There is no a priori reason to expect that the two potential parameters, $\alpha_s$ and $1/a$, should have the same values for four-quark states as in two-quark states. Fortunately, the lattice QCD studies of Ref.~\cite{Okiharu:2004ve} find consistency in this approach for a large set of four-quark spatial configurations. Therefore, we will use fitted values of $\alpha_s$ and $1/a$ from the two-body quarkonia spectrum to calculate the solution to the four-body Schr\"odinger equation and obtain the ground state energy.

\section{Numerical Calculations based on Diffusion Monte Carlo}
\label{sec:numerical}

There are a plethora of ways to calculate the energy levels of many-body non-relativistic systems. For the four-lepton di-positronium molecule, variational methods with a very large amount of trial functions have been used to accurately obtain the ground state energy~\cite{Hylleraas:1947zza,Ho:1986zz,Frolov:1996kq}. Minimizing the Hamiltonian for so many variational parameters, however, can be extremely computationally expensive (see Ref.~\cite{Vijande:2007ix} for a recent attempt). For our numerical calculation, we will instead adopt the more efficient but perhaps less accurate Diffusion Monte Carlo (DMC) method (see Ref.~\cite{Kosztin:1996fh} for an introduction). To test the numerical calculation, we successfully reproduced the binding energy for the di-positronium molecule.

The central idea behind the DMC method is to replace real time by an imaginary time and adjust the guessed ground state energy based on the behavior of the wave function. The time-dependent wave function evolves as $\sum_n e^{-i E_n t} \Psi_n (\vec{x})$, where $E_n$ and $\Psi_n (\vec{x})$ are the true energy eigenvalues and eigenfunctions of the Hamiltonian, respectively. Scaling the energy eigenvalues $E_n$ by a constant guessed value $E_g$ makes no physical change to wave function. Thus, after making the substitutions $E_n \rightarrow E_n - E_g$ and $t \rightarrow - i \tau$, the evolution of the wave function becomes $\sum_n e^{-(E_n - E_g)\tau} \Psi_n (\vec{x})$. When $E_g \approxeq E_0$, only the ground state wave function will be stable while the excited states will diffuse away. If $E_g > E_0$, the wave function will diverge, while for $E_g < E_0$ even the ground state will diffuse away. By adjusting the value of $E_g$ based on the behavior of the wave function, we are able to obtain the correct ground state energy.

Practically, the wave function in the DMC algorithm is represented by random walks of many particles in phase space. To observe the behavior of the wave function with respect to $E_g$, a ``birth-death'' mechanism is implemented such that when $E_g$ is too large, the particles will replicate themselves to increase the total number of particles, and vise versa. For each step, the algorithm adjusts $E_g$ based on the change in the total number of particles until it converges to the ground state energy $E_0$. The stabilized walker distribution then gives us the ground state wave function.

For the system at hand, we also need to find $L_{\rm min}$ for the butterfly configuration. This is similar to the well-known Fermat and Torricelli problem to link three points with a minimal network, which is a special case of the Steiner tree problem in mathematics. The added connecting points in the middle are known as Steiner points, and the original points are called terminals. In two-dimensional space, the Steiner tree problem has an analytical solution, however it is still an NP-complete problem in higher dimensions. To find the positions of the two Steiner points the Steiner configuration for the two vertices  in the right panel of Fig.~\ref{fig:flip-flop-butterfly}, we adopt Smith's algorithm~\cite{smith1992find}. This iterative algorithm determines an equation for each Steiner point, $k$, from summing all possible links surrounding it to both terminals and other Steiner points: $\sum_{kj-{\rm linked}} (\vec{x}^{(i+1)}_k - \vec{x}^{(i+1)}_j)/| \vec{x}^{(i)}_k - \vec{x}^{(i)}_j | = 0$. Here, $\vec{x}^{(i)}_j$ is the position of point $j$ after the $i$-th iteration. Usually after only 20 iterations this algorithm gives a solution that matches the true solution to a very high precision.

\section{Binding Energy for Spin-independent Potential}
\label{sec:SI}
To calculate the tetraquark mass, we will adopt two sets of benchmark parameters for $m_b$, $\alpha_s$ and $1/a$
\beqa
&&\hspace{-3mm}\text{BM-I}:\; m_b = 4.79~\mbox{GeV}\,, \alpha_s = 0.38\,, a = 2.43~\mbox{GeV}^{-1} \,, \nonumber  \\
&&\hspace{-3mm}\text{BM-II}:\; m_b = 5.17~\mbox{GeV}\,, \alpha_s = 0.36\,,  a = 2.34~\mbox{GeV}^{-1} \,. 
\label{eq:benchmark}
\eeqa
Both of them can provide a good fit to the bottomonium spectra~\cite{Quigg:1979vr,Kwong:1987mj}, although the second one requires a universal shift of energy levels of around $-0.77$~GeV, to take into account that the dynamic bottom-quark mass could be different from the bare heavy quark mass. 

\begin{figure}[thb!]
\begin{center}
\includegraphics[scale=0.48]{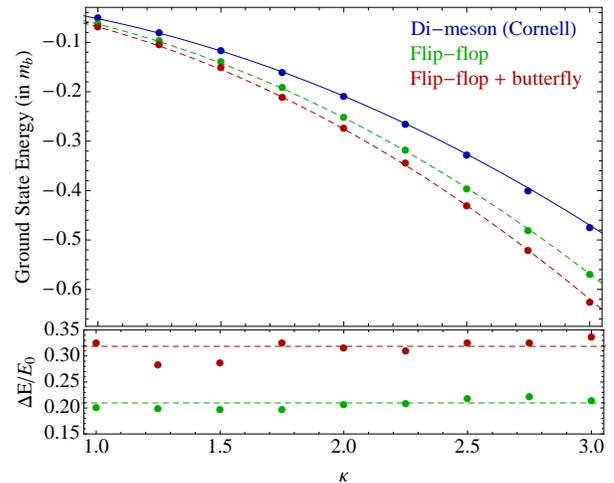} 
\caption{The ground-state energy for three different static potentials as a function of an effective numerical running parameter. The lower panel shows the ratios of binding energy difference for the flip-flop and flip-flop+butterfly configurations over the dissociated di-meson binding energy. Benchmark-I was used for this plot. 
}
\label{fig:cornell-delta-ratio}
\end{center}
\end{figure}
To check the stability of our numerical calculation, we fix the DMC parameters---initial particle number, time step, and simulation length---and modify the total potential by an overall factor of $\kappa$. This is equivalent to scaling the reduced mass and ground state energy while keeping  numerical running conditions fixed.  In dimensionless units, the binding energy is anticipated to scale as $\kappa^2$, which is clear from Fig.~\ref{fig:cornell-delta-ratio}. Furthermore, the ratio of additional binding energy, $\Delta E$, over the di-meson binding energy, $E_0$, for the flip-flop and flip-flop+butterfly potentials should be independent of $\kappa$, which is also approximately true up to small numerical fluctuations. From the lower panel of Fig.~\ref{fig:cornell-delta-ratio}, it is clear that the flip-flop and flip-flop+butterfly configurations account for an additional 20\% and 30\% binding energy, respectively.

\begin{table}[bht!]
\renewcommand{\arraystretch}{1.2}
\begin{center}
  \begin{tabular}{| c | c | c|}
  \hline \hline
  & Benchmark-I   & Benchmark-II  \\ \hline
$E_0/2$ (di-meson)  & 9.455~GeV  & 9.460~GeV \\
       \hline
$\Delta E$ (flip-flop)   & $-52$~MeV   & $-51$~MeV    \\ \hline
$\Delta E$ (flip-flop+butterfly)   & $-80$~MeV   & $-79$~MeV  \\ \hline \hline
  \end{tabular}
  \end{center}
    \caption{The additional binding energy for the two benchmark points.}
  \label{tab:energy-level}
  \end{table}

In Table~\ref{tab:energy-level}, we show the additional binding energy due to the different four-particle effective potential configurations. Within the error of our calculation, we have found that an additional $\sim 50$~MeV and $\sim 80$~MeV of binding energy can be attributed to the flip-flop or flip-flop+butterfly configurations. The total binding energy for the flip-flop+butterfly configuration is around 330~MeV for BM-I and 300~MeV for BM-II (without accounting for the constant energy shift of $-0.77 \times 2$~GeV). These binding energies are large enough to suggest that the four $b$ state should be treated as a true tetraquark system rather than a weakly-coupled molecular system.

\begin{figure}[thb!]
\begin{center}
\includegraphics[scale=0.50]{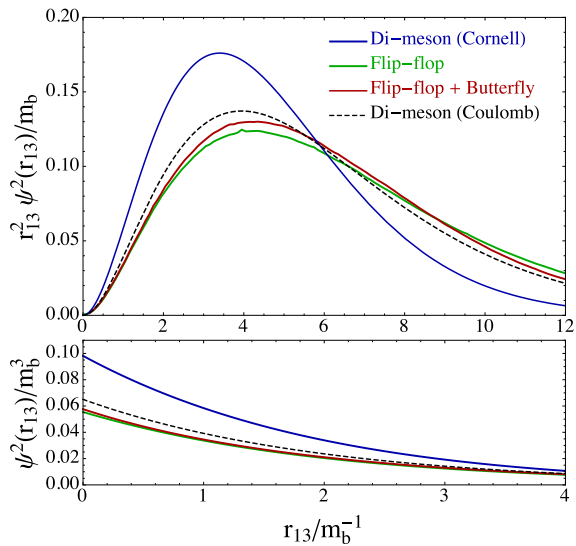} 
\caption{The wave function square in terms of $r_{13}$ for different static potential and for the benchmark fit point-I in Eq.~(\ref{eq:benchmark}).
}
\label{fig:wave-r2psi2}
\end{center}
\end{figure}

Before we move on to discuss the spin-dependent corrections, we show the wave functions for the tetra-quark state. To generate the wave functions in Fig.~\ref{fig:wave-r2psi2} and simplify the multiple-dimension numerical integration, we have treated the wave-function as approximately flat in $r_{12}$ and $r_{34}$ and kept the remaining three variables: $r_{13}$, $r_{14}$ and a relative angle between them. In the upper panel of Fig.~\ref{fig:wave-r2psi2}, we show the squared wave function times $r_{13}^2$ for different static potentials, while in the lower panel we show the value of the squared wave function near the origin. One can see that the squared wave function for the flip-flop+butterfly potential is around 0.59 of the value for dissociated di-meson configuration at the origin. There is only around 5\% difference between the flip-flop and flip-flop+butterfly squared wave functions. We also note that because of the symmetries $\bar{b}_3 \leftrightarrow \bar{b}_4$ and $b_1 \leftrightarrow b_2$, the wave functions are identical when plotted in terms of $r_{14}$, $r_{23}$ or $r_{24}$.

\section{Spin-dependent Corrections}
\label{sec:SD}
For a two particle system, short-range spin-dependent interactions contain a local $\delta$-function force~\cite{DeRujula:1975qlm,Eichten:1980mw}
\beqa
H_{\rm SD} \supset \Delta C_2\,\alpha_s\,\frac{2}{3\,m_1\,m_2}\, \vec{s}_1 \cdot \vec{s}_2 \, 4\pi\,\delta(r_{12}) \,,
\label{eq:SD-general}
\eeqa
with $\Delta C_2$ as the difference of quadratic Casimir. 
For the four particle system, we anticipate the ground state spatial wave function to be predominantly $S$-wave so additional corrections proportional to $\vec{L} \cdot \vec{S}$ should be negligible. From Fig.~\ref{fig:wave-r2psi2}, it is apparent that the wave function for the flip-flop configuration closely approximates the wave function for the flip-flop+butterfly configuration. Thus, to implement the SD correction for the $0^{++}$ tetraquark, we will focus on the flip-flop configuration.

The general ground state wave function for the flip-flop di-meson configuration is
\beqa
\renewcommand{\arraystretch}{1.5}
\hspace{-1.5mm}\Psi = \Bigg\{ 
\begin{array}{ll}
\hspace{-1mm}\psi(r_{13})\,\psi(r_{24})\otimes |(b_1 \bar{b}_3)_1 (b_2 \bar{b}_4)_1 \rangle \otimes \chi^{(0,0)}_{13, 24}     & \;\mbox{for}\,    R_1\,,\\
\hspace{-1mm}\psi(r_{14})\,\psi(r_{23})\otimes |(b_1 \bar{b}_4)_1 (b_2 \bar{b}_3)_1 \rangle  \otimes  \chi^{(0,0)}_{14, 23}   & \; \mbox{for}\,   R_2 \,.
\end{array}
\eeqa
Here, $R_1$ represents the region with $V^{\rm di-meson}_{(13, 24)} < V^{\rm di-meson}_{(14, 23)}$ and otherwise for $R_2$; $(b_1 \bar{b}_3)_1$ represents the color-singlet contraction of $b_1$ and $\bar{b}_3$. The product of the spatial and color wave functions is symmetric under the interchange of $b_1 \leftrightarrow b_2$ and $\bar{b}_3 \leftrightarrow \bar{b}_4$, so the spin-zero wave functions should be
\beqa
\chi^{(0,0)}_{13, 24} &=& \frac{1}{2}(b_1^\uparrow b_3^\downarrow - b_1^\downarrow b_3^\uparrow)(b_2^\uparrow b_4^\downarrow - b_2^\downarrow b_4^\uparrow)   \,, \\
\chi^{(0,0)}_{14, 23} &=& -\frac{1}{2}(b_1^\uparrow b_4^\downarrow - b_1^\downarrow b_4^\uparrow)(b_2^\uparrow b_3^\downarrow - b_2^\downarrow b_3^\uparrow)   \, .
\eeqa
The relative minus between the above two terms is necessary to satisfy the Pauli exclusion principle, and provides the lowest ground state energy after hyperfine splitting.

Calculating the matrix element $\langle \Psi | H_{\rm SD} | \Psi \rangle$, the spin-dependent correction for the $0^{++}$ ground state is approximately
\beqa
\Delta E_{\rm SD} &=& -\,\frac{4\,\alpha_s(\mu)}{3}\,\frac{1}{m_b^2} \, \left[ \psi^2(r_{13} = 0) +  \psi^2(r_{24} = 0) \right]  \nonumber \\
&\approx & -145\pm 30~\mbox{MeV} \,,
\eeqa
for BM-I with $\alpha_s(2\, m_b) \approx 0.2$. The result for BM-II is similar. The symmetries $b_1 \leftrightarrow b_2$ and $\bar{b}_3 \leftrightarrow \bar{b}_4$ imply that the contributions of $R_1$ and $R_2$ to the matrix element are the same, and thus we double the expression for region $R_1$ in our calculation. Here, the relative error is taken to be ${\cal O}(\alpha_s) \approx 20\%$~\cite{Dine:1978yj} and should only be used as guidance from the theoretical calculation. Altogether, the energy for the ground state $0^{++}$ mode is
\beqa
M(0^{++})  = 18.69\pm 0.03~\mbox{GeV}\, ,
\eeqa
which is below the energy threshold of $2M(\eta_b) = 18.798$~GeV and $2M[\Upsilon(1S)] = 18.920$~GeV.

\section{Discussion and Conclusions}
\label{sec:conclusions}

For the di-positronium molecule, the leading decay channel comes from $e^+ e^-$ annihilation, giving $\mbox{Ps}_2 \rightarrow 2 \gamma + e^+ e^-$~\cite{Frolov:1995zz}. Similarly, the leading decay channel for the four-$b$ tetraquark ground state is $0^{++} \rightarrow 2g + b\,\bar{b}$. The decay width is ${\cal O}(10~\mbox{MeV})$ and is comparable to the decay width of $\eta_b$. It can also decay into one on-shell and one off-shell $\Upsilon(1S)$ via spin rearrangement~\cite{DeRujula:1976zlg}, providing a possible four-lepton final state resonance: $0^{++} \rightarrow \Upsilon(1S) \Upsilon(1S)^* \rightarrow \ell^+ \ell^- \ell^+ \ell^-$ with $\ell$ as $e, \mu$.

Within the framework of our calculation, one could also calculate heavier states including spin-one and spin-two excitations. The detailed mass spectrum requires one to calculate the spatial excitation energy as well as the full wave-functions in terms of all degrees of freedom. These are conceptually straightforward, but numerically complicated. Furthermore, one could also apply our calculation procedure to other four-heavy quark system like $c\bar{c}c\bar{c}$ and $c\bar{c}b\bar{b}$. For instance, the ground state, $0^{++}$, of $c\bar{c}c\bar{c}$ is estimated to be below twice of $J/\Psi$ mass and can have the similar decay of $0^{++} \rightarrow \Psi(1S) \Psi(1S)^* \rightarrow \ell^+ \ell^- \ell^+ \ell^-$, which has a smaller branching ratio and different $S/B$ for the experimental searches.

In summary, based on the static potential of the flux-tube model for four heavy quark interactions, we have used a diffusion Monte Carlo algorithm to numerically solve the many-body non-relativistic Schr\"odinger equation. We have found a ground state, $0^{++}$, with a mass of $18.69 \pm 0.03$~GeV, which is approximately 100 MeV below twice the mass of $\eta_b$ expected for a disassociated di-meson ground state. Here, the error of 30 MeV is chosen to include potentially next-to-leading order SD corrections and should not be taken too seriously.

\vspace{3mm}
We thank Joshua Berger, Geoffrey Bodwin, Estia Eichten, and Zhen Liu for useful discussion. This work is supported by the U. S. Department of Energy under the contract DE-FG-02-95ER40896.  


%

\end{document}